\documentclass[aps,prl,reprint,showpacs,groupedaddress]{revtex4-1}

\usepackage{graphicx}
\usepackage{amssymb}
\usepackage{hyperref}
\hypersetup{
    colorlinks=false,
    pdfborder={0 0 0},
}

\begin{document}

\title{Anomalous Behavior of Spin Systems with Dipolar Interactions}

\author{D.~Peter$^1$}
\author{S.~M\"{u}ller$^{2}$}
\author{S.~Wessel$^{3}$}
\author{H.~P.~B\"{u}chler$^1$}
\affiliation{$^1$Institute for Theoretical Physics III, University of Stuttgart, Stuttgart, Germany}
\affiliation{$^2$Max Planck Institute for Physics, 80805 Munich, Germany}
\affiliation{$^3$Institute for Theoretical Solid State Physics, JARA-FIT,  and
JARA-HPC, RWTH Aachen University, Otto-Blumenthal Strasse 26, D-52056
Aachen, Germany}

\date{\today}

\begin{abstract}
 We study the properties of spin systems realized by cold polar molecules interacting via dipole-dipole interactions in two dimensions.
 Using a spin wave theory, that allows for the full treatment of the characteristic long-distance tail of the dipolar interaction, we find several 
 anomalous features in the ground state correlations and the spin wave excitation spectrum, which are absent in their counterparts with short-range interaction.  The most striking consequence is 
 the existence of true long-range order at finite temperature for a two dimensional phase with a broken $U(1)$ symmetry.
\end{abstract}


\maketitle

The foundation for understanding the behavior and properties of quantum matter
is based on models with short-range interactions.  Experimental progress 
in cooling polar molecules \cite{ni08} and atomic gases with large  magnetic dipole moments \cite{griesmaier05}
has  however increased the interest in systems with strong dipole-dipole interactions. While many 
properties of quantum systems with dipole-dipole interactions derive from our understanding 
of systems with short-range interactions,  the dipole-dipole interaction can give rise to phenomena
not present in their short-range counterparts. Prominent examples are the description of dipolar 
Bose-Einstein condensates, where the contribution of the dipolar interaction cannot be included 
in the $s$-wave scattering length \cite{lahaye09}, and the absence of  a first order phase transition with a jump in the density \cite{spivak04}.
In this letter, we demonstrate anomalous  behavior  in two dimensional spin systems with dipolar interactions 
realized by polar molecules in optical lattices.

A remarkable property of cold polar molecules confined into two dimensions  is the potential formation 
of  a crystalline phase for strong dipole-dipole interactions \cite{buechler07,astrakharchik07}.  In contrast to a Wigner crystal with Coulomb 
interactions \cite{bonsall59}, the crystalline phase exhibits the conventional behavior expected for a crystal realized with a
short-range repulsion and the characteristic $1/r^3$ behavior of the dipole interaction can be truncated
at distances involving several interparticle separations. Several strongly correlated phases have been predicted, which
behave in analogy to systems with  interactions extending over a finite range, such as a Haldane phase \cite{torre06}, supersolids  \cite{pollet10,capogrosso10}, pair supersolids in
bilayer systems \cite{trefzger09}, valence bond solids \cite{bonnes10}, 
as well as $p$-wave superfluidity \cite{cooper09}, and self-assembled structures in multilayer setups \cite{wang06}.
On the other hand, it has  recently been demonstrated that polar molecules in optical 
lattices are also suitable for emulating quantum phases of two dimensional spin models  \cite{micheli06,gorshkov11,gorshkov11-2}. 

Here, we demonstrate that such spin models with dipole-dipole interactions exhibit several anomalous features, 
which are not present in their short-range counterparts.  The analysis is based on analytical spin wave theory, which 
allows for the full treatment of the $1/r^3$ tail of the dipole-dipole interactions. We find that the excitation spectrum
exhibits anomalous behavior at low momenta, which gives rise to unconventional dynamic properties of the spin wave excitations.
Remarkably, we derive from this anomalous behavior the existence of a long-range ordered ferromagnetic phase at finite temperatures;
this finding is consistent with the well-known Mermin-Wagner theorem as the  latter does not exclude order
for interactions with a $1/r^\alpha$ tail, where $\alpha \le 4$ \cite{mermin66,bruno01,sousa05}. Finally, we show that the dipole-dipole interaction gives rise to
algebraic correlations even in gapped ground states, in agreement with recent predictions   \cite{deng05,schuch06}.

\begin{figure}[ht]
 \includegraphics[width= 1\columnwidth]{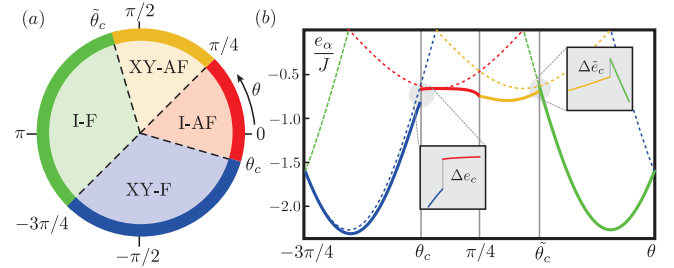}
  \caption{(a) Mean-field phase diagram for the XXZ model  with dipolar interactions, where $\tan \theta$ is the ratio between the XY and the Ising spin couplings. (b) Ground state energy per particle: 
  the dashed lines show the mean-field predictions, 
  while the solid lines include the contributions from the spin waves. At the critical 
  values $\theta_{c}$ and $\tilde{\theta}_{c}$,  the ground state energy exhibits the jump $\Delta e_{c}\approx 0.14J$  and $\Delta \tilde{e}_{c}\approx 0.06J$, 
  indicating the potential formation of an intermediate phase. }\label{fig1}
\end{figure}

We focus on a set up of polar molecules confined into two dimensions in a square lattice,
with each lattice site filled by one polar molecule.  A static electric field applied along 
the $z$ direction splits the rotation levels, and allows us to define a spin 1/2 system by selecting
two states in the rotational manifold. Then, the Hamiltonian reduces to a XXZ model with dipole-dipole interaction
between the spins \cite{gorshkov11}
\begin{equation}
  H = \frac{ J a^3}{ \hbar^2}\sum_{i \neq j} \frac{\cos \theta \: 
S^{z}_{i}S^{z}_{j}+ \sin \theta
\left( S^{x}_{i} S^{x}_{j} + S^{y}_{i} S^{y}_{j}\right) }{|{\bf R}_{i} - {\bf R}_{j}|^3}.
\end{equation} 
Here, the first term accounts for the static dipole-dipole 
interaction between the different rotational levels with strength $J \cos \theta$ , while the last term 
describes the virtual exchange of a microwave photon between the two polar molecules with 
strength $J \sin \theta$, and  $a$ denotes the lattice spacing. 
The dependence of the couplings $J$ and $\theta$ on the microscopic 
parameters is discussed in Ref.~\cite{muller10,gorshkov11,gorshkov11-2} and the one-dimensional version of this model has recently been studied in Ref.~\cite{hauke10}.

Before analyzing this spin model on the square lattice, we present a
summary of the phase diagram
for its counterpart with nearest neighbor interactions only.
Then, the phase diagram is highly symmetric and
exhibits four different phases: (i) an Ising
antiferromagnetic phase (I-AF) for $-\pi/4 < \theta < \pi/4$ with an
excitation gap, (ii) an XY antiferromagnetic phase (XY-AF) for $\pi/4 < \theta
< 3\pi/4$ with a linear excitation spectrum, (iii) an Ising ferromagnetic phase
(I-F) for $3 \pi/4 < \theta < 5\pi/4$ with an excitation gap, and finally (iv) a
XY ferromagnetic phase (XY-F) for $5 \pi/4 < \theta < 7\pi/4$  with a linear
excitation spectrum.

Next, we analyze the modifications of the phase diagram due to dipole-dipole 
interactions between the spins within mean-field theory. The
main influence is the reduction of the stability for the antiferromagnetic
phases, as the next nearest neighbor interaction introduces a  weak frustration
to the system. The ground state energy per lattice site within mean-field
reduces to $  e_{\mbox{\tiny I-AF}} = J \cos \theta \: \epsilon_{\bf K}/4$ and $e_{\mbox{\tiny XY-AF}} = J \sin \theta \: \epsilon_{\bf K}/4$
for the antiferromagnetic phases. The summation over the dipole interaction
reduces to a dimensionless parameter $\epsilon_{\bf K} \approx -2.646$, which is related to
the dipolar dispersion 
\begin{equation}
 \epsilon_{\bf q} = \sum_{j \neq 0} e^{i {\bf R}_{j} {\bf q}}\frac{a^3}{|{\bf R}_{j}|^3}
   \label{dipoledispersion}
\end{equation}
at the corner of the Brillouin zone ${\bf K} = (\pi/a, \pi /a)$.
In turn, the ferromagnetic phases are enhanced with a mean-field energy
$e_{\mbox{\tiny I-F}} =  J \cos \theta \: \epsilon_0/4 $  and $ e_{\mbox{\tiny XY-F}} =  J \sin \theta \: \epsilon_0/4$
with $\epsilon_{0}  \approx 9.033$. The modifications to the
phase diagram are shown in Fig.~\ref{fig1}: first, the Heisenberg points at
$\theta = \pi/4,  5 \pi/4$ are protected by the SU(2) symmetry and still
provide the transition between the Ising and the XY phases. However, the
transitions from the ferromagnetic towards the antiferromagnetic phase are
shifted to the values  $\theta_{c}= \arctan(\epsilon_{\bf K}/\epsilon_{0}) 
\approx -0.1 \pi$ and $\tilde{\theta}_c = \pi+ \arctan(\epsilon_{0}/\epsilon_{\bf K})\approx 0.6 \pi$.

\begin{table*}[t]
\begin{tabular}{c c  c}
\toprule
 Ground state $\alpha$&  \hspace{30pt}  Spin wave excitation spectrum $E^{\alpha}_{\bf q}$   \hspace{30pt} & \hspace{20pt} Ground state energy per spin $e_{\alpha}$ \hspace{20pt} \\
 \hline
\\[-0.2cm]
I-F& $J (\sin \theta  \epsilon_{\bf q} - \cos \theta  \epsilon_{0}) $&    $\displaystyle   \frac{3 J \cos \theta \epsilon_{0}}{4}  +\frac{1}{2}  \int \frac{d{\bf q}}{v_0} E_{\alpha}({\bf q})=\frac{J  \cos \theta \epsilon_{0}}{4}$\\
\\[-0.3cm]
 XY-F & $    J\sqrt{ \sin\theta (\epsilon_{\bf q}-\epsilon_{0})( \cos\theta \epsilon_{\bf q}- \sin \theta \epsilon_{0})}$
&$\displaystyle   \frac{3 J \sin \theta \epsilon_{0} }{4}  +\frac{1}{2}  \int \frac{d{\bf q}}{v_0} E^{\alpha}_{\bf q}$\\
\\[-0.3cm]
 I-AF &  $ J  \sqrt{( \sin \theta \epsilon_{{\bf q}+{\bf K}}-\cos \theta \epsilon_{\bf K}) (\sin \theta \epsilon_{\bf q}- \cos \theta \epsilon_{\bf K })}$
 &$ \displaystyle \frac{3  J \cos \theta \epsilon_{\bf K} }{4}  +\frac{1}{2}  \int \frac{d{\bf q}}{v_0} E^{\alpha}_{\bf q}$\\
\\[-0.3cm]
 XY-AF& $  J  \sqrt{\sin \theta (  \epsilon_{{\bf q}+{\bf K}}-\epsilon_{\bf K} )( \cos \theta \epsilon_{\bf q}-\sin \theta \epsilon_{\bf K})}$& 
 $\displaystyle \frac{3 J \sin \theta \epsilon_{\bf K} }{4} + \frac{1}{2}  \int \frac{d{\bf q}}{v_0} E^{\alpha}_{\bf q}$\\ 
\end{tabular}
\caption{ \label{table1} Spin wave excitation spectrum $E_{\bf q}^{\alpha}$ and ground state energy  $e_{\alpha}$.}
\end{table*}

\begin{figure}[ht]
 \includegraphics[width= 1\columnwidth]{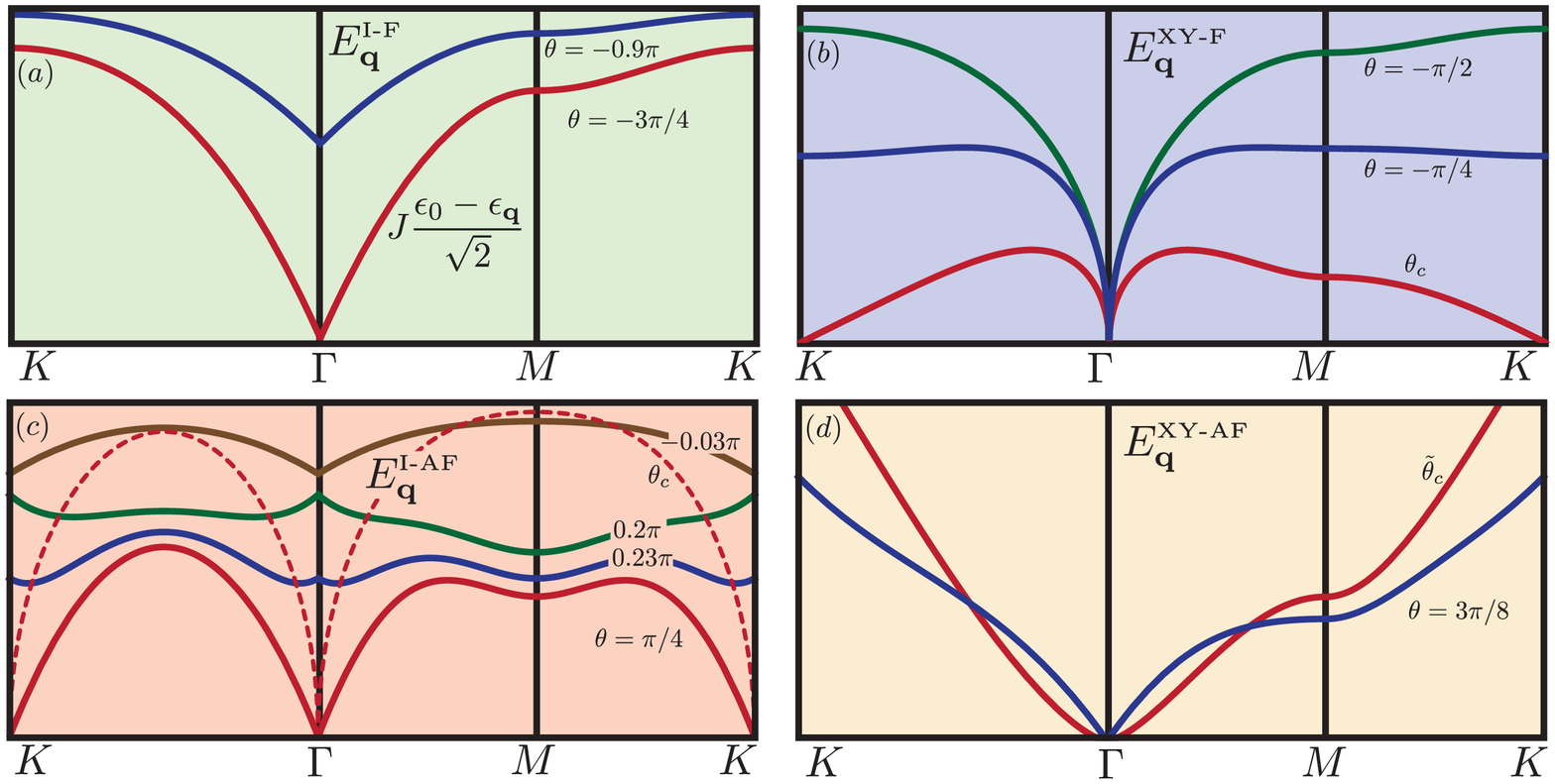}
  \caption{Spin wave excitations with $\Gamma=(0,0)$, $M=(0,\pi/2)$, and $K=(\pi/a,\pi/a)$ for different $\theta$ angles. (a) Spectrum of the I-F phase which also shows the behavior of the dipolar dispersion $\epsilon_{\bf q}$ for $\theta=-3\pi/4$, see red line. 
(b-d) Spectrum for the XY-F, I-AF and XY-AF phases. Each red line is a critical excitation spectrum indicating an instability.
  }\label{fig2}
\end{figure}

The dipole dispersion $\epsilon_{\bf q}$ in Eq.~(\ref{dipoledispersion})
converges very slowly due to the characteristic power law decay of the dipole-dipole
interaction. It is this slow decay, which will give rise to several peculiar
properties of the system. Therefore, we continue first with a detailed discussion
of this dipolar dispersion.  The precise determination of $\epsilon_{\bf q}$ is
most conveniently performed using an Ewald summation \cite{bonsall59}, which transforms
the summation over the  slowly converging terms with algebraic decay into a summation of exponential factors, i.e.,
\begin{eqnarray}
   \epsilon_{\bf q}  & = & - 2 \pi a | {\bf q}|\,{\rm erfc}(a |{\bf q}|/2\sqrt{\pi})   + 4 \pi \left( e^{- \frac{a^2 |{\bf q}|^2}{4 \pi}} - \frac{1}{3}\right)      \label{Ewaldsummation} 
\\
   & & \hspace{-20pt}+     2 \pi \sum_{i\neq 0} \int_{1}^{\infty} \! \frac{d \lambda}{ \lambda^{3/2}} \left[ e^{- \pi  \lambda \left(\frac{ {\bf R}_{i}}{a} + \frac{a {\bf q}}{2 \pi}\right)^2} +\lambda^{2}  e^{- \frac{\pi \lambda |{\bf R}_{i}|^2}{a^2}+ i {\bf R}_{i}{\bf q}}\right] \nonumber
   \end{eqnarray} 
with ${\rm erfc}(x)$ the complementary error function. The important feature of the dipole
dispersion is captured by the first term in Eq.~(\ref{Ewaldsummation}), which
gives rise to a linear and nonanalytic behavior $\epsilon_{\bf q} \sim
\epsilon_{0} - 2 \pi a |{\bf q}|$ for small values $q \ll 1/a $, while all
remaining terms are analytic. It is this linear part, which will give rise to
several unconventional properties of spin systems in 2D with dipolar
interactions, and is a consequence of the slow decay of the dipole-dipole
interaction. The summation in the last term converges very quickly and
guarantees the periodicity of the dipolar dispersion. The quantitative behavior
is shown in Fig.~\ref{fig2}a, and the numerical efficient determination provides
$\epsilon_{0}\approx 9.033$, and $\epsilon_{\bf K}=(1/\sqrt{2}-1)  \epsilon_{0}
\approx -2.646$

Next, we analyze the excitation spectrum above the mean-field ground states
within a spin wave analysis. The spin wave analysis is well
established \cite{kubo52,auerbachbook}, and its application for a spin system with dipolar
interaction is straightforward. Details of the calculation are presented in
the supplementary material for the antiferromagnetic XY phase. The results are
summarized in Table~\ref{table1}, and shown in Fig.~\ref{fig2}. In the following, we present a detailed
discussion for each of the four ordered phases.

{\it Ising ferromagnetic phase:} 
The ferromagnetic mean-field ground state is twofold degenerate with all spins
either point up or down, and is the exact ground state for $\theta =	
\pi$, i.e., $|G\rangle = \prod_{i} \left|\downarrow \right\rangle_{i}$. Within the spin wave analysis,
the ground state is not modified and the excitation
spectrum reduces to $E_{{\bf q}}^{\mbox{\tiny I-F}}$, see Table \ref{table1}.
The spin waves exhibit an excitation gap $\Delta$: (i) approaching the
Heisenberg point at  $\theta = -3 \pi /4$, the excitation gap 
vanishes, indicating the instability towards the XY ferromagnet, (ii) in
turn, for antiferromagnetic XY couplings, the gap is minimal at  ${\bf K}$,
vanishes at the mean-field transition point $\tilde{\theta}_{c}$ and drives an instability towards the formation
of antiferromagnetic ordering.

\begin{figure}[ht]
 \includegraphics[width= 1\columnwidth]{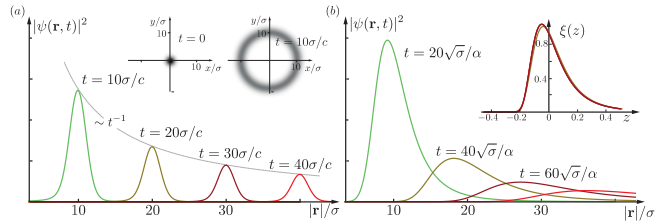}
  \caption{ Time evolution for localized spin excitations described by the Gaussian wave packet $\psi_{0}({\bf r}) = e^{- |{\bf r}|^2/2\sigma^2}/\sqrt{\pi \sigma^2}$ 
  with $\sigma \gg a$ in the continuum description. (a) For a linear dispersion $c|{\bf q}|$ in the I-F phase, the dynamics is described by cylindrical symmetric wave packets (see inset) traveling 
  with velocity $c$, instead of the conventional quantum mechanical spreading for massive systems. 
  (b)  For an anomalous dispersion with $\alpha \sqrt{|{\bf q}|}$ in the XY-F phase, the behavior at long times $t\gg\sqrt{\sigma} \alpha$ reduces to a
  scaling function $\xi(z)$  via   $|\psi(x,\tau)|^2 = \xi(x/\tau -1/2)/\tau^2$ (see inset) using rescaled 
  time $\tau = t \alpha /\sqrt{\sigma}$ and space $x = |{\bf r}|/\sigma$ coordinates. It describes a cylindrical symmetric wave front with velocity 
  $\alpha \sqrt{\sigma}$.}\label{fig3}
\end{figure}

In contrast to any short-range ferromagnetic spin model, the
dispersion relation  $E_{{\bf q}}^{\mbox{\tiny I-F}}$ is not quadratic for small momenta,
but rather exhibits a linear behavior, i.e., $E_{{\bf q}}^{\mbox{\tiny I-F}}   \sim  E_{0}^{\mbox{\tiny I-F}}+ \hbar c |{\bf q}|$ with
velocity $c=  - 2 \pi a J \sin \theta  /\hbar$, which is a consequence of the dipolar interaction in
the system.  This anomalous behavior strongly influences the dynamics
of the spin waves. 
The dynamical behavior of a single localized spin excitation is shown in Fig.~\ref{fig3}a for a Gaussian initial state. 
In order to probe the linear part in the dispersion relation, the width $\sigma$ 
of the localization is much larger than the lattice spacing $a$, and therefore, the dynamics is
well described by a continuum description. 
Instead of the conventional quantum mechanical spreading, one finds  
a ballistic expansion of a cylindrical wave packet with velocity $c$. 
In addition, the dipole-dipole interaction also strongly influences the correlation function.
Within conventional perturbation theory, we find algebraic correlations $\langle S_{i}^{x} S_{j}^{x}\rangle \sim 1/|{\bf r}|^3$.
This algebraic decay of correlations even in gapped systems is a peculiar property of
spin models with long-range interactions  \cite{deng05,schuch06}.

 {\it XY-ferromagnetic phase:} Here, the spins are aligned in the $xy$ plane.
 Within the spin wave analysis, we obtain the excitation spectrum 
 $ E^{\mbox{\tiny{XY-F}}}_{\bf q}$ and the modified ground state energy 
 $e_{\mbox{\tiny{XY-F}}}$.
  In the low momentum regime, the dispersion relation behaves as $ E^{\mbox{\tiny XY-F}}_{\bf q} \sim \sqrt{|{\bf q}|}$,   in contrast to the well 
known  linear Goldstone modes for the broken $U(1)$ symmetry. 
This anomalous behavior is a peculiar property of the dipolar interaction, and the most crucial consequence
is the existence of long-range order for the continuous broken symmetry at finite temperatures 
even in two dimensions \cite{bruno01}.
This property follows immediately from the above spin wave analysis:
the order parameter reduces to  $m \equiv \Delta m - 1/2=\langle S_{i}^{x}\rangle/\hbar $, where $\Delta m$  accounts for 
the suppression of the order parameter by quantum fluctuations. Within spin wave theory, it reduces to ($\Delta m = \langle a^{\dag}_{i} a_{i }\rangle$)
\begin{displaymath}
\Delta m\! = \!\!  \int \frac{d{\bf q}}{v_{0}}\left[\frac{\cos \theta \epsilon_{\bf q} +\sin \theta ( \epsilon_{\bf q}-2 \epsilon_{0})}{4 E_{\bf q} } \coth \left(\frac{E_{\bf q}}{2T}\right) - \frac{1}{2}\right] .
\end{displaymath}
This expression is finite and small: at $T=0$, the integrand behaves as  $\sim 1/\sqrt{|{\bf q}|}$ and we find a 
suppression of the order $\Delta m \approx 0.008$ at $\theta= - \pi/2 $. The smallness of this corrections due to
quantum fluctuations is a good justification for the validity of the spin wave analysis.  On the other hand, 
even at finite temperatures, the low momentum behavior of the integrand takes the form $\sim T/|{\bf q}|$, 
and provides a finite contribution in contrast to a conventional Goldstone mode, which provides a logarithmic divergence.

The appearance of a long-range order at a finite temperature for a ground state with a broken $U(1)$ symmetry is
a peculiar feature of dipole-dipole interactions, which renders the system more mean-field like.  The system therefore
exhibits a finite temperature transition at a critical temperature $T_{c}$ into a disordered phase; such a behavior is 
consistent with the classical XY model with dipolar interactions \cite{bruno01}. The correlation functions determined
within spin wave theory and a high temperature expansion are summarized in Table \ref{table2}.
 Note, that the spin wave analysis neglects the influence of vortices. This is well justified here, 
as the dipolar interactions give rise to a  confining of vortices, i.e., the interaction potential 
between a vortex--antivortex pair increases linearly with the separation between the vortices.

The spin wave dynamics caused by the anomalous dispersion relation $\sim \sqrt{|{\bf q}|}$ is shown in Fig.~\ref{fig3}b for a Gaussian wave packet of width $\sigma$. Interestingly, the propagation velocity of the wave packets is proportional to $\sqrt{\sigma}$ and thus faster for broad wave packets, in contrast to the usual dispersion dynamics. This is a consequence of the group velocity $v_{\bf q} \sim 1/\sqrt{|{\bf q}|}$ which is large for the small momentum components involved in the broad wave packets. 

\begin{table}
\begin{tabular}{c c c c}
\toprule
 correlation function &  \hspace{10pt} $T=0$    \hspace{10pt}&   \hspace{10pt} $0 < T< T_{c}$    \hspace{10pt}&   \hspace{10pt}$T_{c}< T $    \hspace{10pt}\\
 \hline
$ \langle S^{z}_{i} S^{z}_{j}\rangle$  &$ \sim |{\bf r}|^{-5/2}$  & $\sim |{\bf r}|^{-3}$ &  $\sim |{\bf r}|^{-3}$ \\
$ \langle S^{y}_{i} S^{y}_{j}+ S^{x}_{i} S^{x}_{j}\rangle- m^2$  &$ \sim |{\bf r}|^{-3/2}$ &  $\sim |{\bf r}|^{-1}$ &  $\sim |{\bf r}|^{-3}$
\end{tabular}
\caption{ \label{table2} Correlation functions in the XY-F phase predicted by the spin wave analysis and high temperature expansion. }
\end{table}

{\it Ising antiferromagnetic phase}: Next, we focus on the antiferromagnetic phases and start with the I-AF ground state.  
Again, the ground state is twofold degenerate on bipartite lattices.  We choose the ground state with spin up on sublattice $A$
and spin down on sublattice $B$, i.e., $|G\rangle = \prod_{i \in A} \left|\uparrow\right\rangle_{i} \prod_{j\in B} \left|\downarrow\right\rangle_{j}$. 
The spin wave analysis is straightforward (see supplementary information), and we obtain
the spin wave excitation spectrum $ E^{\mbox{\tiny I-AF}}_{\bf q}$ and ground state energy $ e_{\mbox{\tiny I-AF}}$, see Table \ref{table1}.
The system exhibits an excitation gap as expected for a system with a broken $Z_{2}$ symmetry. 
However, the dipole interactions give rise to an anomalous behavior at small momenta similar to 
the ferromagnetic Ising phase with  $E^{\mbox{\tiny I-AF}}_{\bf q}- E^{\mbox{\tiny I-AF}}_{0} \sim -
\sin \theta |{\bf q}|$.  Consequently, the dynamics of spin waves at low momenta is analogous to the 
Ising ferromagnet,  see Fig.~\ref{fig3}.  Within spin wave theory, we obtain that the antiferromagnetic correlations  
$\langle (-1)^{i-j} S_{i}^{\beta}S_{j}^{\beta} \rangle$ 
and the ferromagnetic correlations  $\langle S_{i}^{\beta}S_{j}^{\beta} \rangle$ decay with the  
power law  $\sim1/ |{\bf r}|^{3}$ with $\beta = x,y,z$ determined by the characteristic behavior of the dipole-dipole interaction. 
The excitation gap vanishes approaching the mean-field critical point $\theta_{c}$ towards the XY- ferromagnetic 
phase, and also while approaching the Heisenberg point at $\theta = \pi/4$. For the latter, the qualitative behavior of the excitation spectrum 
changes drastically within a very narrow range of $\theta$, see Fig.~\ref{fig2}c.

{\it XY antiferromagnetic phase}:  Finally, we analyze  the properties of the antiferromagnetic XY phase.
In contrast to the ferromagnetic XY phase, the excitation spectrum  $E^{\mbox{\tiny XY-AF}}_{\bf q}$ 
exhibits the conventional linear Goldstone mode, see Fig.~\ref{fig2}d. This can be understood, as the 
antiferromagnetic ordering introduces a cancellation of the dipolar interactions, and provides a behavior in analogy to
its short-range counter part: true long-range order exists only at $T=0$, while at finite temperature the system
exhibits quasi long-range order and eventually undergoes a Kosterlitz-Thouless transition for increasing temperature.
Nevertheless, the dipole-dipole interactions give rise to the characteristic algebraic correlations, e.g., 
$\langle (-1)^{i-j}S_{i}^{z}S_{j}^{z} \rangle \sim 1/|{\bf r}|^3$ for the antiferromagnetic transverse spin correlation at zero temperature.

Finally, we comment on the transitions between the different phases. The spin wave analysis predicts, 
that the excitation spectrum for each phase becomes unstable at the mean-field critical points: For the Heisenberg 
points at $\theta = \pi/4, 5\pi/4$, such a behavior is expected due to the enhanced symmetry and one indeed finds, 
that at the critical point, the excitation spectrum from the Ising phase coincides with the spectrum from the XY ground state.   
Consequently, the spin waves provide the same  contribution to the ground state energy, see Fig.~\ref{fig1}b. In turn, at the instability points 
$\theta_{c}$ and $\tilde{\theta}_{c}$, the excitation spectrum from the antiferromagnetic phase is different from the spectrum for the ferromagnetic $F$ phase.
Consequently, the ground state energy within the spin wave analysis exhibits a jump, see Fig.~\ref{fig1}a. Such a behavior is an indication 
for the appearance of an intermediate phase. However,  this question cannot be conclusively answered within the presented analysis due to  the limited validity of spin wave theory close to the transition points. However, the appearance of a first order phase transition can be excluded by arguments similar to
Ref.~\cite{spivak04}.

 \begin{acknowledgments}
We thank M. Hermele for helpful discussions. Support by the Deutsche Forschungsgemeinschaft (DFG) within SFB
/ TRR 21 and National Science Foundation under Grant No. NSF PHY05-51164
is acknowledged.
 \end{acknowledgments}

\newpage

\section*{Supplementary information}
We present the derivation of the spin wave excitation spectrum
within the spin wave analysis. The basic approach is to start with the ground
states exhibiting perfect order, which are the correct ground states for the classical 
model at the four points $\theta = 0,\pm \pi/2, \pi$.  Then, we introduce bosonic
creation and annihilation operators creating a spin excitation above the ground
state according to the Holstein-Primakoff transformation. The spin Hamiltonian
then reduces to a Bose-Hubbard model. In lowest order, we can ignore the
interactions between the bosonic particles, and obtain a quadratic Hamiltonian
in the bosonic operators, which is diagonalized using a
Bogoliubov-Valantin transformation.  The latter transformation deforms the
ground state and introduces fluctuations into the system. 

In the following, we demonstrate the spin wave analysis for the most revealing case: the antiferromagnetic XY phase. 
The generalization to the other ground states is straightforward.  Without loss of generality, we choose the antiferromagnetic order 
to point along the  $x$ direction. The square lattice is bipartite, and we denote the two sublattices as A and B. Then, the antiferromagnetic
 mean-field ground-state is given by   $|G\rangle = \prod_{i\in A} \left|\leftarrow\right\rangle_{i} \prod_{j\in B} \left|\rightarrow\right\rangle_{j}$ 
 with spins on sublattice~A pointing in the negative $x$ direction, i.e., $S^x \left|\leftarrow\right\rangle = -\hbar/2 \left|\leftarrow\right\rangle$, 
 and spins on sublattice~B pointing in the positive $x$ direction ($S^x \left|\rightarrow\right\rangle = \hbar/2 \left|\rightarrow\right\rangle$).
 Excitations on sublattice A are created by flipping a spin with the ladder operator $S^{x+} = S^z-i S^y$, 
 while excitations on sublattice B are created via $S^{x-} = S^z+i S^y$. 
 We apply a Holstein-Primakoff transformation to bosonic operators 
\begin{eqnarray} \label{transHolstein}
S^z_i = \frac{\hbar}{2} (a^{\phantom\dag}_i + a^{\dag}_i) \,\varphi(n_i), \hspace{15pt}
S^y_i = \frac{\hbar}{2i} (a^{\phantom\dag}_i - a^{\dag}_i)\, e^{i {\bf K} {\bf R_i}} \,\varphi(n_i), \nonumber
\end{eqnarray}
where the phase $e^{i {\bf K}{\bf R_i}}=e^{-i {\bf K}{\bf R_i}}$ accounts for the sublattice-dependent sign with ${\bf K} = (\pi/a, \pi/a)$. The factor $\varphi(n_i)=1-n_i$ is introduced to guarantee bosonic commutation relations for the operators $a_i$.  Here, we are interested in the leading order of the spin wave expansion, and can therefore set 
 $\varphi(n_i) \approx 1$. The bosonic operators reduce to
\begin{eqnarray}
       a_i = (S^z +i S^y e^{i {\bf K} {\bf R_i}} )/\hbar,  \hspace{20pt}   a^\dag_i = (S^z -i S^y e^{i {\bf K} {\bf R_i}})/\hbar, \nonumber
\end{eqnarray}
and the number operator $n_i = a^\dag_i a^{\phantom \dag}_i = \frac{1}{2} + S_x e^{i {\bf K}{\bf R_i}}/\hbar$.

Expanding the spin Hamiltonian in terms of the bosonic operators leads to a Bose-Hubbard Hamiltonian for the spin wave excitations. 
In leading order, we can neglect the interactions between the bosons and obtain
\begin{eqnarray} \label{HamiltonianXYAF}
\frac{H}{J}& = &\sin \theta \, \epsilon_{\bf K} \left(\frac{3 N}{4} - \frac{1}{2}\sum_i\left[a^{\dag}_{i}a^{\phantom\dag}_{i}+ a^{\phantom\dag}_{i} a^{\dag}_{i}\right]\right) \\
 & &\hspace{-30pt}+  \frac{1}{4} \sum_{i\ne j} \frac{
\chi_{i j}\left(a^{\dag}_{i} a^{\phantom\dag}_{j}  + a^{\phantom\dag}_{i} a_{j}^{\dag}\right) 
 +\eta_{i j}\left(a^{\phantom\dag}_{i} a^{\phantom\dag}_{j}  + a^{\dag}_{i} a^{\dag}_{j} \right)
}{|{\bf R}_{ij}/a|^3} \nonumber
\end{eqnarray}
with  ${\bf R}_{ij}={\bf R}_i - {\bf R}_j$,  $N$ the number of lattice sites, and the coupling the terms
$\chi_{i j}=\cos\theta+\sin\theta e^{i {\bf K} {\bf R}_{ij}}$ and $\eta_{i j}=\cos\theta-\sin\theta e^{i {\bf K} {\bf R}_{ij}}$ including the
 antiferromagnetic ordering.
Introducing the  Fourier representation $a_i =  \sum_{\bf q} a^{\phantom\dag}_{\bf q}e^{-i {\bf q} {\bf R}_i}/\sqrt{N}$,
the terms involving the bosonic operators in Eq.~(\ref{HamiltonianXYAF}) reduce to
\begin{eqnarray}
\frac{1}{4} \sum_{\bf q} \bigg[
\left( \cos\theta \, \epsilon_{\bf q} +\sin\theta \, \epsilon_{\bf q+K} - 2 \sin \theta \epsilon_{\bf K}\right)\Big(a^{\dag}_{\bf q} a^{\phantom\dag}_{\bf q}  + a^{\phantom\dag}_{\bf q} a_{\bf q}^{\dag}\Big)  \nonumber\\
+\left( \cos\theta \, \epsilon_{\bf q}-\sin\theta \, \epsilon_{\bf q+K} \right)\Big(a^{\phantom\dag}_{\bf q} a^{\phantom\dag}_{-\bf q}  
+ a^{\dag}_{\bf q} a^{\dag}_{-{\bf q}} \Big) \bigg].  \nonumber
\end{eqnarray}
The diagonalization of this Hamiltonian is straightforward using a standard Bogoliubov transformation
with $b^{\dag}_{\bf q} = u_{\bf q}\, a^{\dag}_{\bf q} - v_{\bf q}\, a^{\phantom\dag}_{-\bf q}$.
Then, the Hamiltonian takes the form 
\begin{eqnarray} \label{Hdiagonal}
H = \frac{3J N \sin \theta\, \epsilon_{\bf K}}{4}  + \sum_{\bf q} E^{\mbox{\tiny XY-AF}}_{\bf q} \left(b^{\dag}_{\bf q}b^{\phantom\dag}_{\bf q} + \frac{1}{2}\right)
\end{eqnarray}
with the spin-wave excitation spectrum $E^{\mbox{\tiny XY-AF}}_{\bf q}$. In addition, the coefficients  for the Bogoliubov transformation 
are given by
\begin{eqnarray}
\!\!u_{\bf q} , v_{\bf q} = \pm\sqrt{\frac{1}{2} \left( \frac{\cos\theta \, \epsilon_{\bf q} +\sin\theta \, (\epsilon_{\bf q+K} - 2 \epsilon_{\bf K})}{2\, \mathcal{E}_{\bf q}}\pm 1\right)}
\end{eqnarray}
with $\mathcal{E}_{\bf q} \equiv E^{\mbox{\tiny XY-AF}}_{\bf q} / J $.
The property $u^2_{\bf q}-v^2_{\bf q}=1$ asserts that the transformation is canonical. 
In addition, the ground state obeys the condition $b_{\bf q} \left|\text{vac}\right\rangle = 0$, and the ground state energy per spin at zero temperature $T=0$
reduces to $e_{\mbox{\tiny XY-AF}}$, see Table~\ref{table1}.

We are now able to check the validity of the spin wave approach self-consistently: the deformation of the ground state by the spin wave analysis
provides a suppression of the antiferromagnetic order $ m \equiv  \Delta m - \frac{1}{2}=\left\langle S^x_i e^{i {\bf K} {\bf R_i}} \right\rangle / \hbar$,
and thus
\begin{eqnarray}
\Delta m= \int \frac{d{\bf q}}{v_0} \, \big\langle a^{\dag}_{\bf q} a^{\phantom\dag}_{\bf q} \big\rangle = \int \frac{d{\bf q}}{v_0} \Big[ v_{\bf q}^2 + (2 v_{\bf q}^2 + 1) f_{\bf q} \Big],
\end{eqnarray}
where $f_{\bf q} = \big\langle b^{\dag}_{\bf q}b^{\phantom\dag}_{\bf q} \big\rangle = \left[\exp(E^{\mbox{\tiny XY-AF}}_{\bf q}/ T)-1\right]^{-1}$ accounts for the thermal occupation of the spin waves. At zero temperature $T=0$, this expression converges and  we obtain $\Delta m \approx 0.03$ for $\theta = \tilde\theta_c$ as well as $\Delta m = 0.39$ for $\theta \approx \frac{\pi}{4}$. In turn, at finite temperatures $T>0$, the low momentum behavior of  the integrand scales as $|{\bf q}|^{-2}$, and therefore $\Delta m$ diverges logarithmically: the long range order is destroyed by the thermal spin wave fluctuations, and gives rise to the well-known 
quasi long-range order in analogy to short-range XY models.

Finally, the spin wave analysis also allows us to analyze the correlation functions
 $c_{\alpha\alpha}({\bf R}_{ij})=\langle S^\alpha_i S^\alpha_{j} e^{i {\bf K} {\bf R}_{i j}} \rangle$. Using
 the translational invariance of our system, the correlation functions reduce to 
\begin{eqnarray} \label{correlationFunction}
c_{\alpha\alpha}({\bf r}) = \int \frac{d{\bf q}}{v_0} \, c_{\alpha\alpha}({\bf q+K}) \, e^{-i {\bf q} {\bf r}}  
\end{eqnarray}
with $c_{\alpha\alpha}({\bf q}) = \langle S^\alpha_{\bf q} S^\alpha_{\bf -q}\rangle$. 
Combining the Holstein Primakoff transformation  Eq.~(\ref{transHolstein}) and the Bogoliubov transformation 
the correlation functions can be expanded in terms of the coefficients $u_{\bf q}$ and $v_{\bf q}$
\begin{eqnarray}
c_{zz}({\bf q+K}) &=& \frac{1}{4} (u_{\bf q+K}+v_{\bf q+K})^2 \coth\left(\frac{J \mathcal{E}_{\bf q}}{2T}\right) , \nonumber\\
c_{yy}({\bf q+K}) &=& \frac{1}{4}  (u_{\bf q}-v_{\bf q})^2 \coth\left(\frac{J \mathcal{E}_{\bf q}}{2T}\right) .
\end{eqnarray}
The long distance behavior $|{\bf r}| \rightarrow \infty$ of the correlation function is determined by the low momentum 
behavior of the above expressions
\begin{eqnarray}
(u_{\bf q+K}+v_{\bf q+K})^2 &\sim& |{\bf q}|+\text{const.} \nonumber\\
(u_{\bf q}-v_{\bf q})^2 &\sim&\frac{ 1}{|{\bf q}|}
\end{eqnarray}
and describes the leading non-analytic part. 
The latter can be replaced using the following relation, which derives via an Ewald summation,
\begin{equation}
   |{\bf q}|^{\gamma} \sim \sum_{j \neq 0}\frac{e^{i {\bf q} {\bf R}_{j}}}{|{\bf R}_{j}|^{2+\gamma}}
\end{equation}
for $|{\bf q}|\rightarrow 0$ and $\gamma >-2$; (for $\gamma = 0,2,4,\ldots$ the left side is replaced by $|{\bf q}|^\gamma \log |{\bf q}|$). 
At zero temperature $T=0$, the integration in Eq.~(\ref{correlationFunction})
is straightforward and provides the scaling behavior  $c_{zz}({\bf r}) \sim |{\bf r}|^{-3}$ and $c_{yy}({\bf r}) \sim |{\bf r}|^{-1}$.


\begin{thebibliography}{10}

\bibitem{ni08}
K.-K. Ni et~al.,
\newblock Science {\bf 322}, 231 (2008).

\bibitem{griesmaier05}
A.~Griesmaier, J.~Werner, S.~Hensler, J.~Stuhler, and T.~Pfau,
\newblock Phys. Rev. Lett. {\bf 94}, 160401 (2005).

\bibitem{lahaye09}
T.~Lahaye, C.~Menotti, L.~Santos, M.~Lewenstein, and T.~Pfau,
\newblock Reports on Progress in Physics {\bf 72}, 126401 (2009).

\bibitem{spivak04}
B.~Spivak and S.~A. Kivelson,
\newblock Phys. Rev. B {\bf 70}, 155114 (2004).

\bibitem{buechler07}
H.~P. B{\"u}chler et~al.,
\newblock Phys. Rev. Lett. {\bf 98}, 060404 (2007).

\bibitem{astrakharchik07}
G.~E. Astrakharchik, J.~Boronat, I.~L. Kurbakov, and Y.~E. Lozovik,
\newblock Phys. Rev. Lett. {\bf 98}, 060405 (2007).

\bibitem{bonsall59}
L.~Bonsall and A.~A. Maradudin,
\newblock Phys. Rev. B {\bf 15}, 1959 (1977).

\bibitem{torre06}
E.~G. Dalla~Torre, E.~Berg, and E.~Altman,
\newblock Phys. Rev. Lett. {\bf 97}, 260401 (2006).

\bibitem{pollet10}
L.~Pollet, J.~D. Picon, H.~P. B\"uchler, and M.~Troyer,
\newblock Phys. Rev. Lett. {\bf 104}, 125302 (2010).

\bibitem{capogrosso10}
B.~Capogrosso-Sansone, C.~Trefzger, M.~Lewenstein, P.~Zoller, and G.~Pupillo,
\newblock Phys. Rev. Lett. {\bf 104}, 125301 (2010).

\bibitem{trefzger09}
C.~Trefzger, C.~Menotti, and M.~Lewenstein,
\newblock Phys. Rev. Lett. {\bf 103}, 035304 (2009).

\bibitem{bonnes10}
L.~Bonnes, H.~P. B\"uchler, and S.~Wessel,
\newblock New Journal of Physics {\bf 12}, 053027 (2010).

\bibitem{cooper09}
N.~R. Cooper and G.~V. Shlyapnikov,
\newblock Phys. Rev. Lett. {\bf 103}, 155302 (2009).

\bibitem{wang06}
D.-W. Wang, M.~D. Lukin, and E.~Demler,
\newblock Phys. Rev. Lett. {\bf 97}, 180413 (2006).

\bibitem{micheli06}
A.~Micheli, G.~K. Brennen, and P.~Zoller,
\newblock Nature Physics {\bf 2}, 341 (2006).

\bibitem{gorshkov11}
A.~V. Gorshkov et~al.,
\newblock Phys. Rev. Lett. {\bf 107}, 115301 (2011).

\bibitem{gorshkov11-2}
A.~V. Gorshkov et~al.,
\newblock Phys. Rev. A {\bf 84}, 033619 (2011).

\bibitem{mermin66}
N.D.~ Mermin and H.~Wagner,
\newblock Phys. Rev. Lett. {\bf 17}, 1133 (1966).

\bibitem{bruno01}
P.~Bruno,
\newblock Phys. Rev. Lett. {\bf 87}, 137203 (2001).

\bibitem{sousa05}
J.~R. de~Sousa,
\newblock Eur. Phys. J. B {\bf 43}, 93 (2005).

\bibitem{deng05}
X.-L. Deng, D.~Porras, and J.~I. Cirac,
\newblock Phys. Rev. A {\bf 72}, 063407 (2005).

\bibitem{schuch06}
N.~Schuch, J.~I. Cirac, and M.~M. Wolf,
\newblock Commun. Math. Phys. {\bf 267}, 65 (2006).

\bibitem{muller10}
S.~ M\"uller, {\it Diploma thesis}, University of Stuttgart (2010).

\bibitem{hauke10}
P.~Hauke, F.~M.~Cucchietti, A.~M{\"u}ller-Hermes, M.~Bañuls, and J.~Ignacio,
\newblock New Journal of Physics {\bf 12}, 113037 (2010).

\bibitem{kubo52}
R.~Kubo,
\newblock Phys. Rev. {\bf 87}, 568 (1952).

\bibitem{auerbachbook}
A.~Auerbach,
\newblock {\em Interacting Electrons and Quantum Magnetism},
\newblock Springer-Verlag, New York, 1994.

\end{thebibliography}
\end{document}